\documentclass[acmsmall, authordraft]{acmart}
\AtBeginDocument{%
  \providecommand\BibTeX{{%
    \normalfont B\kern-0.5em{\scshape i\kern-0.25em b}\kern-0.8em\TeX}}}

\setcopyright{acmcopyright}
\copyrightyear{2021}
\acmYear{2021}
\acmDOI{10.1145/1122445.1122456}

\acmJournal{TiiS}

\usepackage{color}
\usepackage{colortbl, booktabs}

\usepackage{xcolor}
\usepackage{multirow}
\usepackage{rotating}
\usepackage{subcaption}

\begin{document}

\title{Towards Involving End-users in Interactive Human-in-the-loop AI Fairness}

\renewcommand{\shorttitle}{Involving End-users in Human-in-the-loop AI Fairness}

\author{Yuri Nakao}
\authornote{Contact author}
\email{Nakao.Yuri@fujitsu.com}
\orcid{0000-0002-6813-9952}
\affiliation{%
  \institution{Research Center for AI Ethics, Fujitsu Limited}
  \city{Kawasaki}
  \country{Japan}
}

\author{Simone Stumpf}
\email{Simone.Stumpf@glasgow.ac.uk}
\orcid{0000-0001-6482-1973}
\affiliation{%
  \institution{University of Glasgow}
  \city{Glasgow}
  \country{UK}
}

\author{Subeida Ahmed}
\email{Subeida.Ahmed@city.ac.uk}
\affiliation{%
  \institution{City, University of London}
  \city{London}
  \country{UK}
}

\author{Aisha Naseer}
\email{Aisha.Naseer@uk.fujitsu.com}
\affiliation{%
  \institution{Fujitsu Research of Europe Ltd.}
  \city{Hayes}
  \country{UK}
}

\author{Lorenzo Strappelli}
\email{Lorenzo.Strappelli@city.ac.uk}
\affiliation{%
  \institution{City, University of London}
  \city{London}
  \country{UK}
}

\renewcommand{\shortauthors}{Nakao, Stumpf, et al.}

\begin{abstract}
Ensuring fairness in artificial intelligence (AI) is important to counteract bias and discrimination in far-reaching applications. Recent work has started to investigate how humans judge fairness and how to support machine learning (ML) experts in making their AI models fairer. Drawing inspiration from an Explainable AI (XAI) approach called \emph{explanatory debugging} used in interactive machine learning, our work explores designing interpretable and interactive human-in-the-loop interfaces that allow ordinary end-users without any technical or domain background to identify potential fairness issues and possibly fix them in the context of loan decisions.  Through workshops with end-users, we co-designed and implemented a prototype system that allowed end-users to see why predictions were made, and then to change weights on features to "debug" fairness issues. We evaluated the use of this prototype system through an online study. To investigate the implications of diverse human values about fairness around the globe, we also explored how cultural dimensions might play a role in using this prototype. Our results contribute to the design of interfaces to allow end-users to be involved in judging and addressing AI fairness through a human-in-the-loop approach. 
\end{abstract}



\maketitle

\section{Introduction}
Many organizations are now using artificial intelligence (AI) systems to support their decision making, for example, in clinical decision support systems \cite{CaiHCMLMedical} and in bail decisions \cite{Mallari2020.DoILookLikeACriminal}. However, these AI systems might be biased and therefore decisions might be unfair, or even discriminatory \cite{kirkpatrick_battling_2016}.
A common focus has been to make AI decisions fairer by relying on machine learning (ML) experts. Many toolkits and systems have been developed to help ML experts to investigate fairness \cite{AdebayoFairML, Bllamy2019AIF360, friedler_comparative_2019, Wexler2019WhatifTool, Cabrera-fairvis-2019, ahn_fairsight:_2019, yan_silva_2020, galhotra_fairness_2017, holstein_improving_2019, aggarwal_black_2019}, often through quantitative fairness measures, and then removing fairness issues through algorithmic means, for example, by re-balancing the training data set or changing weights on attributes \cite{friedler_comparative_2019, aggarwal_black_2019, galhotra_fairness_2017, Barocas2016.BigDataDI}. However, approaching fairness only through quantitative means is challenging as it is difficult to choose the appropriate metric to use for a specific context, and the qualitative and value-driven aspects of human fairness perceptions.

Relying only on ML experts ignores the perspectives that end-users can bring to AI fairness. Recent work has explored what end-users consider fair and how this could be expressed in fairness metrics \cite{Srivastava_Mathematical_2019}. Alongside this strand of research exploring human perceptions of fairness in decision-making, the need to directly involve end-users in assessing the fairness of AI systems and making ML models fairer has recently been emphasized \cite{binns_its_2018, dodge_explaining_2019,holstein_improving_2019, veale_fairness_2018, yan_silva_2020, Yu_DIS20}. There is already some work that aims to include end-users in the AI development process through value sensitive design practices \cite{Kirkham_DIS20, Ballard_DIS19},  technology demonstrators \cite{Wouters_DIS19}, or through participatory design practices \cite{lee2019.WebuildAI}. However, our work aims to build interpretable and interactive human-in-the-loop tools that allow end-users without any technical or domain background to assess the fairness of an AI system, and to find and fix issues if necessary.

Our research draws inspiration from Explainable AI (XAI) research to make AI models interpretable~\cite{doshi-velez_towards_2017}. We are particularly focused on making these systems work for end-users without a background in fairness, AI or domain knowledge. Our work specifically uses \emph{explanatory debugging for interactive machine learning \cite{kulesza_principles_2015}} as a lens through which to approach human-in-the-loop fairness tools. Explanatory debugging aims to help end-users to identify and correct "bugs" (i.e. mispredictions that fail to meet the user’s expectations) in an AI model.  The approach includes a tightly coupled cycle of interactions with a user interface (UI) to explain to the end-user how the ML model makes its decisions and then provide an opportunity for the user to make necessary corrections. This framework also proposes eight principles for designing UIs. We leverage this existing framework for AI fairness; we provide explanations of how the ML model made its decisions so that end-users can identify and correct fairness issues that do not meet their expectations. We believe that this can provide a fresh viewpoint for the design of interpretable and interactive human-in-the-loop AI fairness tools. 

In this paper we focus on the design a human-the-loop AI system for loan decisions and we investigate how end-users, i.e. people without any special technical or domain knowledge who are the receiving end of loan decisions, use this system to identify potential fairness issues and improve fairness. Because of recent work that has demonstrated the importance of cultural dimensions on human fairness perceptions~\cite{Silvernail2016CrossCulturalPhd, Blake2015NatureOntologyOfFairness, Bolton2010.HowDoPriceFairness, MATTILA2006CrossCulturalComparison, KIM200783FormingandReacting, GM05, LEUNG:2008}, we also investigated potential cultural dimensions in the way that end-users assess fairness and try to fix perceived fairness issues using this human-in-the-loop system, by drawing on Hofstede's Cultural Dimensions (CDs) framework \cite{hofstede_dimensionalizing_2011}. This framework argues that human behavior is shaped by six dimensions that vary across cultures: Power Distance, Individualism, Masculinity, Uncertainty Avoidance, Long Term Orientation, and Indulgence. Our research questions were:
\begin{enumerate}
\item How can we design interactive human-in-the-loop interfaces for end-users to help them identify and fix fairness issues?
\item How do end-users assess whether an AI system is fair or unfair, using a human-in-the-loop interface? 
\item Can these end-users fix fairness issues through their feedback?
\item Are there any cultural dimensions to fairness assessments and use of the prototype?
\end{enumerate}

To answer these research questions, we designed a prototype human-in-the-loop system, inspired by principles of explanatory debugging \cite{kulesza_principles_2015} and drawing on requirements from end-users who took part in a series of workshops in the United States of America (USA), the United Kingdom (UK), and Japan with 12 participants. This was then followed by an online empirical study of 388 participants using the prototype. 

Through our work, we advance the continuing conversation around building tools to investigate fairness and including end-users in the AI development process. In particular, we contribute to:
\begin{enumerate}
     \item a better understanding of how \emph{end-users} assess fairness, and their requirements for tools that support them,
     \item extending our knowledge of how cultural dimensions influence how end-users make fairness assessments and use these tools in making these assessments,
     \item designing new interactive human-in-the-loop interfaces to support end-users in making AI fairness assessments,
    \item increasing end-users ability to be directly involved in shaping AI systems and potentially making them fairer.
\end{enumerate}

The structure of this work is as follows. We first provide an overview of related work in AI fairness, explainable AI (XAI) and explanatory debugging, and cultural dimensions. We then present a prototype that was designed based on co-design workshop findings. Next, we present an empirical online study of the prototype system. We finish the paper with a discussion of the wider implications of our work.

\section{Related Work}
We provide an overview of previous work which has started to investigate what fairness is, how it can be measured objectively, and how humans perceive AI fairness. We describe existing tools which aim to help ML experts to produce fair AI systems, and review work that has started to involve end-users in building fair AI systems. We then outline the explanatory debugging approach, and its relevance to AI fairness. We conclude with details of the cultural dimensions framework, which underlies our research into cultural aspects of AI fairness. 

\subsection{Fairness in Human Decision-making and AI}

Fairness is often associated with group fairness and individual fairness \cite{Binns2020.IVandGFairness}. Group fairness \cite{verma_fairness_2018} means that the same treatment is provided to people belonging to protected groups, defined by sensitive attributes such as race or gender, as to other groups. Individual fairness \cite{Bllamy2019AIF360} focuses on providing the same treatment for individuals having similar or same characteristics or sensitive attributes. 

To determine if something is fair, many quantitative fairness metrics have been proposed \cite{verma_fairness_2018}; for example, independence (e.g. demographic parity or statistical parity); separation (e.g. equality of opportunity, equality of odds, other variations); and sufficiency (e.g. predictive parity, predictive quality). 
Demographic parity is sometimes termed disparate impact in legal contexts, which is the demographic parity of the protected group divided by that of the privileged group. According to US law, if disparate impact is less than 0.8, there is discrimination on a protected attribute \cite{FeldmanKDD2015CerRemDI}.
Currently, demographic parity is considered as a ‘gold’ standard baseline statistical measurement \cite{wachter2020fairness}, however which particular fairness metric to use depends on the use case or the business problem at hand; in short, the metrics need to be applied in context.

Hence, recent work relevant to intelligent user interfaces has studied fairness from the perspective of human judgement, and the values and criteria that are applied by humans in judging fairness. Many of these research efforts have been directed at understanding what humans consider fair, and how individuals’ demographics impact fairness assessments or cause biases.
Recent work found that non-experts' ideas of fairness in the context of criminal risk and skin cancer risk prediction most closely matches demographic parity \cite{Srivastava_Mathematical_2019}. In another study \cite{woodruff_qualitative_2018}, end-users rejected the use of sensitive attributes and the 'stereotyping' of groups of individuals. In the context of financial decisions, there has been some work to involve users in assessing the credit risk, for example, to investigate decision-making \cite{Green2019Principles}, or to examine people's fairness perceptions and definitions \cite{Saxena2019AIES.HowFairDefFare}. One such study found that people tended to give preferential treatment to the people belonging to a protected group \cite{Saxena2019AIES.HowFairDefFare}. This echoes findings from experiments conducted to better understand fairness perceptions  \cite{Wang2020FactInfluPerFair} which showed that algorithms that predict in people’s favour are likely to be rated as fairer despite having bias against particular demographic groups. Lee \cite{Lee_Understanding_2018} measured perceived fairness from people’s perceptions of algorithms in decision making for different tasks (i.e., ones that require mechanical skills and human skills) and revealed that with tasks requiring human skills, algorithmic decisions were perceived as less fair. Mallari et al. \cite{Mallari2020.DoILookLikeACriminal} examined the impact of racial information on human decision making with respect to users' judgements of recidivism and found that race of the defendants (black defendant vs. white defendants) had significant impact on the user’s decision about the defendants.  Overall perception of AI fairness has been found to be related to individuals' computer-based knowledge and privacy concerns
~\cite{araujo2020ai}.

All of above studies highlight the importance of exploring perceived fairness and fairness criteria in decision-making processes. 
In section 3.2, we describe the results of conducting workshops with end-users to explore their criteria and process when assessing fairness, and how we instantiated this in a prototype. Further, in section 4.5, we explore how people consider fairness in the context of loan decisions using this prototype through a online user study. Thus, we aim to extend the discussion of fairness criteria and clarify how the criteria vary depending on cultures, which underpins research in human-in-the-loop fairness. 

\subsection{Algorithms and Interfaces to Support Fairness}

With the proliferation of algorithmic decision-making systems in public and private sectors, it becomes crucial to evaluate and address their social impact, such as bias or discrimination. Thus, a lot of research has been conducted into developing fair decision-making algorithms \cite{Chouldechova2020Snapshot, friedler_comparative_2019, aggarwal_black_2019, galhotra_fairness_2017, Barocas2016.BigDataDI}. 
These approaches often apply quantitative fairness metrics (see section 2.1) that test for and then remove fairness issues, for example, by re-balancing the training data set \cite{friedler_comparative_2019} or setting weights on attributes \cite{aggarwal_black_2019, galhotra_fairness_2017} or even masking bias \cite{Barocas2016.BigDataDI}. 

Several bias mitigation tools including open source libraries have been developed to help ML experts with developing fair AI models, for example, FairML \cite{AdebayoFairML}, AI Fairness 360 \cite{Bllamy2019AIF360}, Fairness comparison \cite{friedler_comparative_2019}, and Google's What-if tool \cite{Wexler2019WhatifTool}. 
In addition to these industrial tools, there are some tools developed in a research context. For example, a visual analytics system, FairSight \cite{ahn_fairsight:_2019}, has been designed to evaluate fairness through understanding, measuring, diagnosing and mitigating biases. 
Silva \cite{yan_silva_2020} is a fairness optimisation tool for exploring potential sources of unfairness in datasets or machine learning models using causality or causal graphs.
Additionally, Cabrera et al.~\cite{Cabrera-fairvis-2019} developed a visual analytics system called FairVis for discovering intersectional bias.
These tools support ML experts (or users familiar with AI) with an organizational stake in fairness to investigate performance, biases, and typically use standard fairness metrics. 

However, these systems are not targeted at \emph{end-users} who are the receiving end of the decisions being made. The need to design and develop systems and interfaces to help end-users identify AI biases has been well-recognised \cite{RobertICIS2019, holstein_improving_2019, veale_fairness_2018}.
Studies have explored the needs from practitioners in industry~\cite{holstein_improving_2019} or public sector~\cite{veale_fairness_2018}. To involve end-users in AI development processes, Lee et al. \cite{lee2019.ProceduralJustice, lee2019.WebuildAI} attempted to enable people to build algorithmic policy for on-demand food donations \cite{lee2019.WebuildAI}, and also gave end-users the opportunity to improve algorithmic fairness through AI interfaces guaranteeing procedural justice \cite{lee2019.ProceduralJustice}. Work has also been conducted to elicit the fairness preferences of stakeholders and then build them into an optimisation algorithm \cite{Zhang_AIES20}. 

Consequently, designing and building interactive human-in-the-loop AI fairness tools for \emph{end-users} is thus far underexplored. To our knowledge, only Silva \cite{yan_silva_2020} has been evaluated with non-ML experts, and it does not allow fixing of fairness issues. The prototype we present in section 3.2 shows how an interface can be designed targeted at \emph{end-users} so that they can investigate potential fairness issues, based on requirements gathered during co-design. Moreover, our prototype allows end-users to feed back to the AI model, in order to fix potential fairness issues. We show the feasibility of this approach in section 4.5.2.

\subsection{Explainable AI (XAI) and Explanatory Debugging}

Our work is aligned with some of the concerns of Explainable AI (XAI) to make AI systems interpretable \cite{doshi-velez_towards_2017}. Providing explanations of ML systems is important for end-users so that they can understand why decisions were made, improve their mental models \cite{kulesza_tell_2012} and to calibrate their trust in the system \cite{gunning_xaiexplainable_2019}. Much work has been focused on how to make the AI model more transparent, for example, by showing important features of individual decisions of the AI model \cite{ribeiro_why_2016}. It has been argued that explanations should be provided to answer users' questions around What, Why, Why Not, How, What-if \cite{lim_why_2009, bellotti_intelligibility_2001} by following a transparency design process \cite{eiband_bringing_2018}. There have also been increasing efforts to design explanations through a user-centered Human-Computer Interaction (HCI) perspective. It has been argued that explanations can be presented in various ways \cite{wang_designing_2019}, such as textual, graphical or hybrid explanations, depending on user expertise \cite{Szymanski_IUI21}. Some work has warned that explanations can be misleading and can influence users to over-rely on AI decisions \cite{bussone_role_2015, Chromik_IUI21}. There has been lots of work on how to evaluate explanations with users, for example in terms of trust and mental model soundness \cite{kulesza_too_2013, hoffman_metrics_2018}, and the choice of tasks that play a role in evaluating AI systems \cite{Bucina_IUI2020}.

Clearly, much of this work also applies to interfaces to investigate fairness, because an AI model has to be interpretable for the user to be able to understand how fair it is. Many of the tools discussed in section 2.2 already use XAI mechanisms and explanations in their design implicitly, however, there are very few research efforts that have investigated \emph{fairness} from an explicit XAI angle. A notable exception is Dodge et al. \cite{dodge_explaining_2019} who examined the role of 'templated' explanations to support end-users in judging the fairness of AI systems, noting that there is no one 'right' explanation style but that both global and local explanations might be necessary to make fairness assessments.

 In our work, we explicitly adopt a user-centered XAI approach. We use \emph{explanatory debugging for interactive machine learning \cite{kulesza_explanatory_2010,  kulesza_principles_2015}} as a lens through which to approach human-in-the-loop fairness tools. Explanatory debugging aims to help end-users to identify and correct 'bugs' (i.e. mispredictions that fail to meet the user’s expectations) in an AI model \cite{stumpf_interacting_2009, kulesza_explanatory_2010, groce_you_2014}. The approach includes a tightly coupled cycle of interactions with a user interface (UI) to explain to the end-user how the ML model makes its decisions \cite{kulesza_tell_2012} and then provide an opportunity for the user to make necessary corrections \cite{kulesza_fixing_2009}. We extend this approach to \emph{explanatory fairness debugging}, where a user aims to identify and correct biases that do not conform to their expectations of fairness (i.e. 'fairness bugs') through a tight cycle of interaction with an interface.

To help design explanatory debugging interfaces, eight principles have been proposed~\cite{kulesza_principles_2015}. Explanations should be 1) iterative, 2) be sound, 3) be complete but 4) do not overwhelm. This means that explanations should be concise, interactive, and in-situ, are truthful to the underlying model and show as much information that the system uses in making predictions, but this needs to be carefully traded off against what users can process and understand. The feedback to the system should 5) be actionable, 6) be reversible, 7) honored, and 8) be shown through incremental changes. This can be achieved by allowing users to make interactive corrections on the explanations provided, undo their corrections, integrating their feedback into the model, and being able to see if their corrections have the desired effect (or going toward the desired effect). We show in section 3.2 how we instantiated our human-in-the-loop fairness interface to follow these explanatory debugging principles.

\subsection{Cultural aspects}
\label{sec:CulturalAspects}
In order to explore cultural differences in fairness assessments, we now discuss approaches for analysing culture.
Culture, viewed as the characteristics, knowledge and behavior of a group of people, has been studied extensively, alongside cross-cultural aspects. These aspects in different cultures have been viewed and analysed in many different ways. For example, cultures can fall along a spectrum of 'high context' and 'low context' cultures depending on the importance of context in interpersonal communication~\cite{hall1976beyond}. Parsons and Shils~\cite{parsons1962toward} discussed differences between cultures in terms of five patterns of behaviors that are reflected at individual levels and social systems levels. Many more frameworks have been proposed but the most influential has been Geert Hoftstede's framework of Cultural Dimensions (CDs).

In the 1970s, Hofstede started to develop a multi-dimensional cultural model, based on research on employees in IBM globally~\cite{hofstede_dimensionalizing_2011}. 
The model consists of six dimensions around which cultures differ \cite{10.1080/13602381003637609, hofstede_dimensionalizing_2011}. These six dimensions are:  
\begin{itemize}
    \item \textbf{Power Distance (PD)} relates to the extent to which unequal power distribution, including inequality, is expected and accepted by the less powerful in society. 
    \item \textbf{Individualism (IDV)} describes the degree to which people are integrated into groups and the strength of the ties within these groups. The other end of this dimension is Collectivism. 
    \item \textbf{Masculinity (MSC)} refers to how assertive and competitive a society is, versus Femininity, which is more modest and caring.
    \item \textbf{Uncertainty Avoidance (UA)} describes a society's tolerance for ambiguity, and unstructured, novel situations.
    \item \textbf{Long Term Orientation (LTO)} refers to societies which like to maintain time-honoured traditions and norms while viewing societal change with suspicion.
    \item \textbf{Indulgence (IDG) } is related to a society's support for  free gratification of basic and natural human drives related to enjoying life and having fun.
\end{itemize}

The framework allows quantitative comparison between cultures based on wide-scale surveys that focus not on the variations in individual responses but instead try to find tendencies at the country level\footnote{https://geerthofstede.com/research-and-vsm/vsm-2013/}. Hofstede makes it clear that the survey questions cannot be used to derive individual scores~\cite{hofstede_dimensionalizing_2011}, as it is not meant as a psychological personality instrument. As a result of applying these surveys over the last 50 years, country-based CD scores have been developed and are readily available\footnote{Scores can be obtained from https://www.hofstede-insights.com/country-comparison/}.

Hofstede's work on Cultural Dimensions has not been without criticism in terms of methodology and conceptualization of culture on a national level \cite{mcsweeney_hofstedes_2002}. It has often been accused of using unsound survey techniques among a limited group of participants, being reductive in terms of cultural differences that span national borders, or that it does not take into account sub-groups within a larger cultural framework. Hofstede \cite{hofstede2002dimensions} has countered that the methodology has been validated across a very large sample of data, by other researchers, spanning more than 50 years, using a variety of research techniques. While he acknowledges that nationality is not the best unit for studying cultures, they are usually the only kind of units available for comparison. In any case, as he argued, dimensions should be seen as "constructs, which have to prove their usefulness by their ability to
explain and predict behavior"  \cite{hofstede2002dimensions} and thus might not be the only way to understand people's behavior.

Despite these criticisms, it is widely used to understand interactions with and design considerations for technology, for example, it has been employed to analyze cultural differences in mobile service design \cite{10.1145/1054972.1055064} and well as security behaviour \cite{10.1145/3025453.3025926}. 
It has also been extensively validated~\cite{hofstede2010cultures} and used to understand how fairness is culturally dependent ~\cite{Silvernail2016CrossCulturalPhd, Blake2015NatureOntologyOfFairness, Bolton2010.HowDoPriceFairness, MATTILA2006CrossCulturalComparison, KIM200783FormingandReacting, GM05, LEUNG:2008}.

In this paper, we employ Hofstede's framework to explore cultural differences in our empirical data, in order to investigate whether they can explain differences in fairness assessments.  Instead of splitting participants into countries or regions in which they are based, we follow Hofstede's suggested approach and map each participant’s CD scores from the country score given by Hofstede framework. We then compare across culture, by investigating differences between high and low CD scores of each dimension, as we explain further in section \ref{sec:CS_DataAnalysis}.

\section{An Interactive Human-in-the-loop AI fairness prototype for End-users}
In order to investigate how we can design interactive human-in-the-loop interfaces for end-users to help them identify and fix fairness "bugs" (RQ1), we obtained requirements from a series of co-design workshops and coupled them with principles of explanatory debugging. We first describe the setup of the workshops, then describe the prototype's functionality and rationale for its design.

\subsection{Co-Design Workshops}
Following a co-design approach to interface design \cite{sanders_co-creation_2008, stumpf_design_2021} , we ran a series of workshops in the USA, UK, and Japan with a total of 12 end-users. We recruited 3 participants (2 women, 1 man, mean age 47.3) for the co-design workshops held in the USA, 5 participants (3 women, 2 men, mean age 34.2) in the UK, and 4 participants (3 women, 1 man, mean age 33.75) in Japan through social media and personal contacts. As end-users, these participants had no special technical or domain knowledge, and we only checked that they were engaged in the topic, by recruiting people who had previously applied for a loan. Since our work is meant to help people explore and assess fairness using their own criteria rather than pre-established fairness metrics, we also did not require previous knowledge or engagement with any fairness research. We paid an incentive of £40, or equivalent in the local currency. The study was approved by the City, University of London Computer Science Research Ethics Committee; all participants were over 18 years old and informed consent was obtained before they took part in the co-design workshops.

For each country, we held 2 co-design workshops; these two workshops were 3 weeks apart and each lasted 2 hours.Due to COVID-19 restrictions, we were unable to conduct face-to-face workshops, and thus held all workshops online. In workshop 1 we mainly investigated how end-users went about looking at fictitious loan application scenarios that we developed, and what information they wanted to identify potential fairness issues. Based on the feedback obtained in workshop 1, we then constructed an initial low-fidelity prototype UI, and evaluated and refined it during workshop 2. The workshops in the USA, and UK were conducted in English and in Japan conducted in Japanese by researchers who are native speakers of these languages.  We recorded all workshops and used thematic analysis \cite{braun_using_2006} to develop code sets around what information or cues they looked for to assess fairness, and how we could design and improve UIs for assessing fairness. 

\begin{figure*}[t]
\centering
\includegraphics[width=\linewidth]{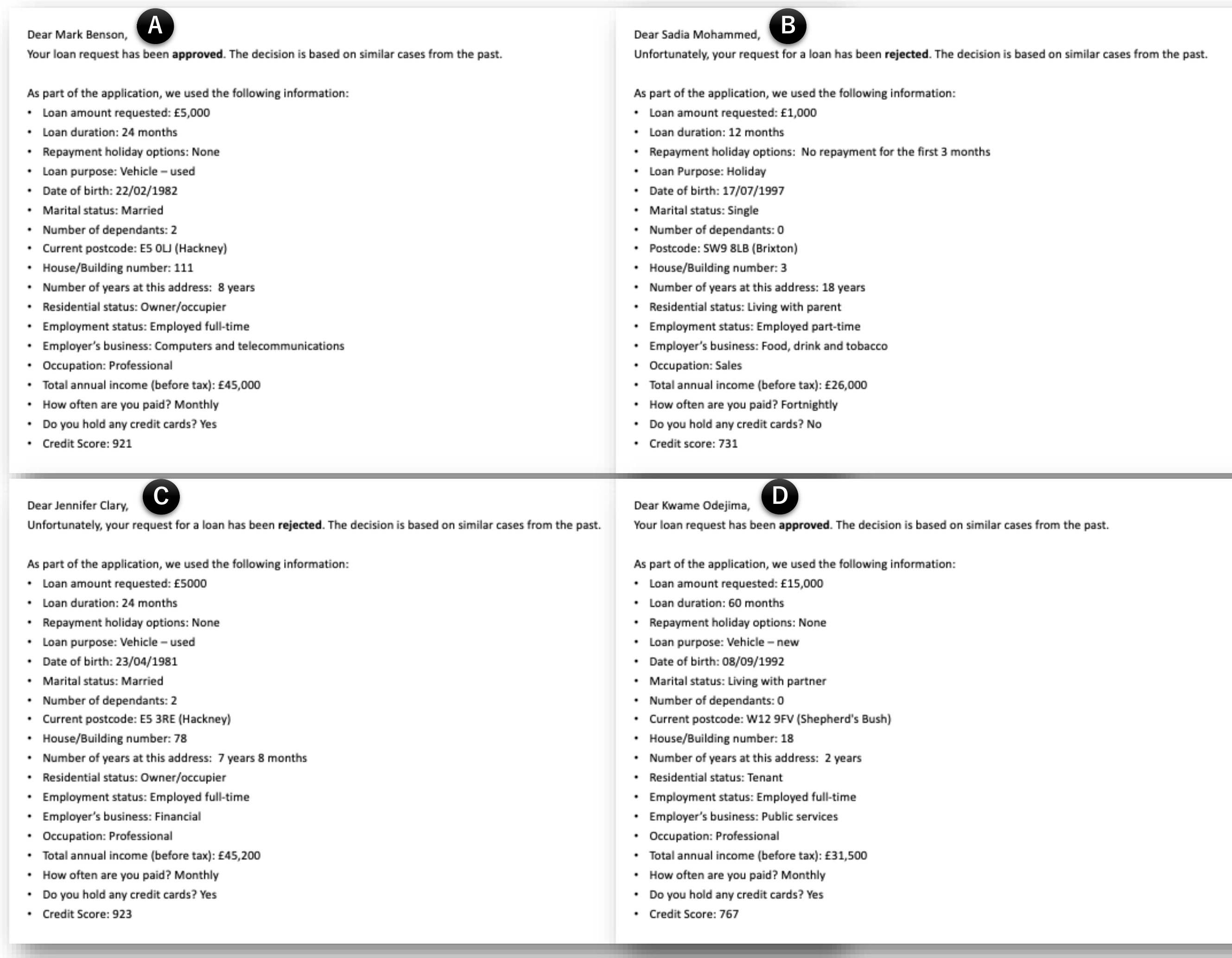}
\caption{(A) Application 1: Mark Benson, (B) Application 2: Sadia Mohammed, (C) Application 3, Jennifer Clary, and (D) Application 4: Kwame Odejima}
\Description[Screenshots (A)-(D) of the four workshop text-based applications]{}
\label{fig:workshop_slides}
\end{figure*}

\subsubsection{Workshop 1 procedure} 
In workshop 1, we explored how end-users explore fairness in loan decisions, using their own criteria. To investigate what attributes and information they might be looking for to assess the fairness of the applications' outcomes decided by AI and potentially what they would change to make the decisions fairer, we developed 4 fictitious applications as scenarios to consider. We changed some of the application scenario details to localize them to each country (e.g. names, currency, dates, addresses) but otherwise kept them the same. The language of the slides is translated by a researcher carefully avoiding changing the meaning. The application details that we showed to end-users are typical of information collected as part of a loan application process, based on the application form of a well-known international bank. Credit scores of applicants are frequently obtained and used by financial institutions to assess the risk of lending applicants money.

Application 1 (USA/UK: Mark Benson or Japan: Kazufumi Takahashi, Fig. \ref{fig:workshop_slides} A) was always approved, as it was a 'safe' application, with a homeowner with a very good credit score applying for a small loan to buy a used car. Application 2 (USA/UK: Sadia Mohammed or Japan: Chihe Pak, Fig. \ref{fig:workshop_slides} B) was rejected, as it was a more 'risky' application with low income, part-time job and low credit score. We also included her application to investigate any potential minority or age biases. Application 3 (USA/UK: Jennifer Clary or Japan: Maika Suzuki, Fig. \ref{fig:workshop_slides} C) was also rejected but crucially her details were very similar to application 1. This was to introduce an application that seemed, without any further information, to be clearly unfair. Finally, application 4 (USA/UK: Kwame Odejima or Japan: D\~{u}ng Nguy\^{e}n, Fig. \ref{fig:workshop_slides} D) was accepted although it seemed more 'risky'. 

Each scenario was discussed in turn within each group. This discussion focused on exposing whether end-users thought the decision was fair and why, based on the information included in the application or their experience of the decisions they had seen, and what information would have been useful for them to assess fairness better.

\subsubsection{Development of an initial prototype} 
\begin{figure}[t]
\centering
\includegraphics[width=\linewidth]{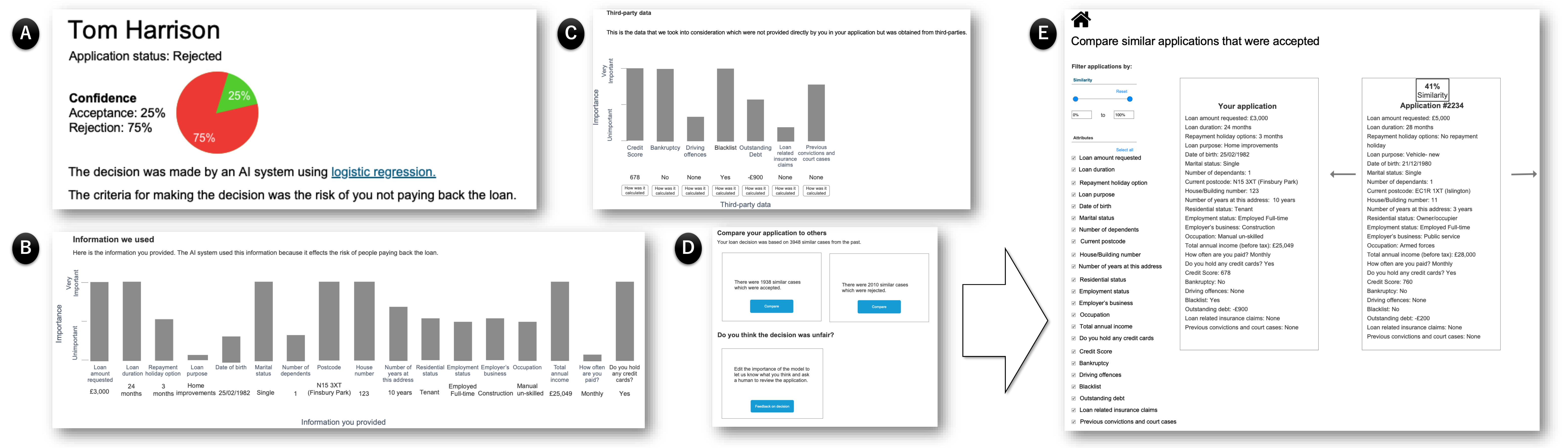}
\caption{The clickable wireframes used in Workshop 2. (A)-(D) are on the same page. By clicking a blue button in (D), users can move to page (E). The details of each component is as follows: (A) The summary of decision, with the predicted decision, confidence, and explanation of an AI model (logistic regression). (B) The values and weights for the attributes on the application such as loan amount requested, purpose of loans or their address. (C) The values and weights for the attributes provided from a third party organization such as credit score, or the history of driving offences. (D) Buttons to transition to a comparison page showing similar rejected or accepted cases. (E) Similar cases and their attribute values.}
\Description[Screenshots (A)-(F) of the prototype GUI used in Workshop 2]{(A) shows the applicant name, application status, the prediction confidence also displayed as a pie chart, and the algorithm and criteria the AI used to make the decision. The values and weights for the attributes that the customer provided to the bank (B) and those that were provided from a third party organisation (C) are displayed as two separate bar charts. (D) shows three call to actions, the first two ask the user to compare the application with other similar applications and the last whether the user thought the decision was fair or unfair. Screen (E) is the destination screen a user would be directed to after clicking a call to action on screen (D).}
\label{fig:home}
\end{figure}

We analyzed workshop 1 using a thematic code set that focused on  information and cues for assessing fairness that were mentioned by participants during the discussion of the scenarios. We then mapped these codes to interface design elements to construct clickable wireframes to instantiate their suggestions in an interface (Fig. \ref{fig:home}). 

\subsubsection{Workshop 2 procedure}In workshop 2, we structured our discussion on the clickable wireframes' screens, using scenarios that explored a number of fictitious loan applications. Going through each screen's functionality, we discussed what helped to understand if the application decisions were considered fair, what additional information would they like to determine fairness, and what feedback they would like to give to fix fairness issues.

We further analyzed the feedback during workshop 2 using thematic analysis \cite{braun_using_2006} to develop and refine requirements for a human-in-the-loop prototype system, which we describe in detail next.

\subsection{The Interactive Human-in-the-loop Prototype System}
\begin{figure}[t]
\centering
\includegraphics[width=\linewidth]{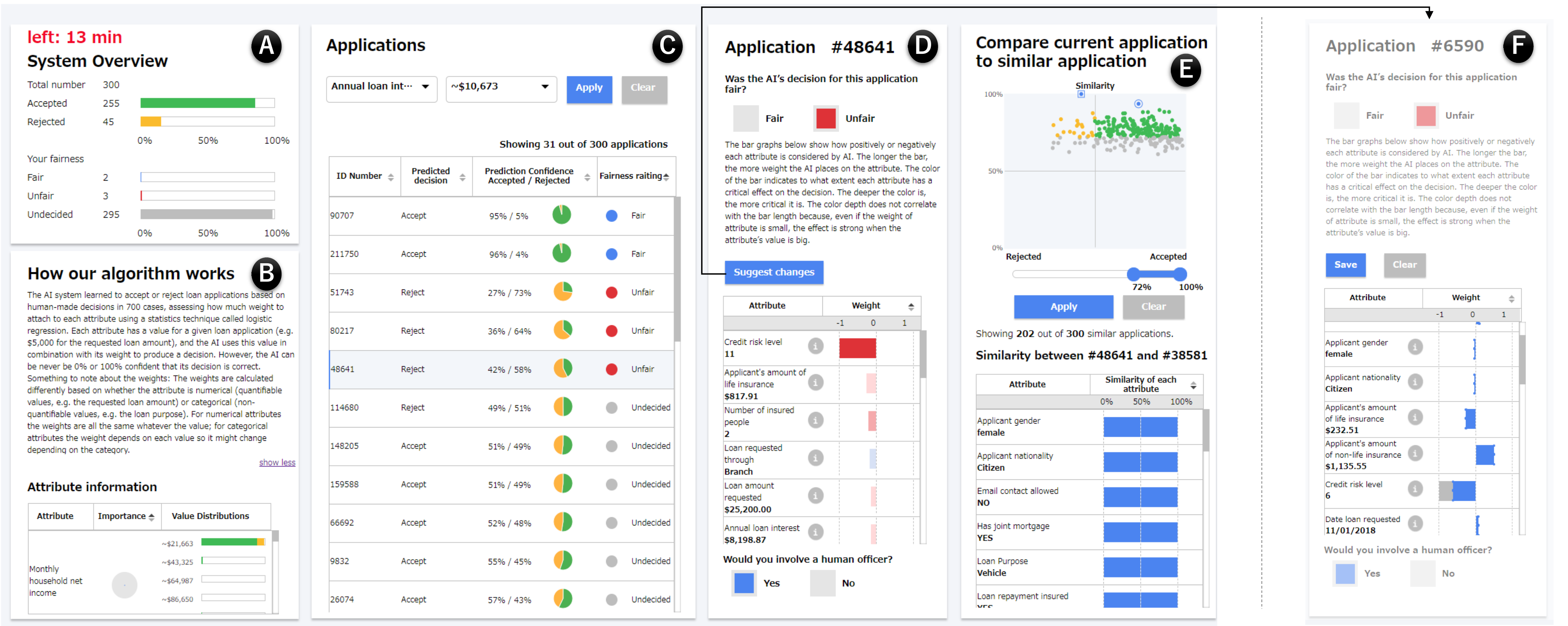}
\caption{The final interface to support fairness assessments.
Section (A) shows how many applications are accepted and rejected by the AI system, and judged fair or unfair by the user.
(B) shows more information about the overall model. It contains an explanation of the algorithm, the models' attributes, their absolute weight values (their ``importance''), and the distribution of the attribute values with respect to the accept/reject decisions.
(C) shows an overview of applications which can be filtered on the attribute values. Each application is presented with whether it's been accepted or rejected by the AI model, the confidence of the decision, and whether it's been marked as fair or unfair. 
(D) shows more details for each applications. The weight for each attribute is shown graphically in a bar chart. The colour saturation represents the criticality of the attribute value on the decision of the application. We also allow users to modify these weights.  Users can mark up the evaluate as fair or unfair, and whether this application should be looked at by a human. 
(E) the scatter plot provides an overview of the currently selected application against all other applications in relation to their similarity (y-axis) and decision confidence (x-axis). The scatter plot can be filtered by similarity. An application can be compared to the currently selected application, showing their similarity for each attribute. (F) by clicking Suggest Changes button in (D), participants can move to Suggest Change mode (F). Participants can adjust the weights for attributes by dragging the bar graphs.}
\Description[Screenshots (A)-(F) of the final interface]{The interface is comprised of visual analytics which include bar charts, pie charts, scatter plots, filters and interactive bars to set weightings. The interface is supported by textual explanations.}
\label{fig:GUI}
\end{figure}

We now describe each panel and its functionality in more detail, justifying its design through requirements arising from our analysis of the workshops (shown in \emph{italics}) and Explanatory Debugging principles relating to providing explanations (shown as \texttt{font}).  While we also looked for patterns in the feedback that might indicate cultural differences in the information or functionality that should be included, we did not find any that affected the overall design.

An overview of the prototype system's interface is shown in Fig.\ref{fig:GUI}. The prototype interface consists of 4 main panels: System-wide information (Fig.\ref{fig:GUI} (A) and (B)), an overview of applications (Fig.\ref{fig:GUI} (C)), more details of a currently selected application (Fig.\ref{fig:GUI} (D)), and information comparing the currently selected application with other applications (Fig.\ref{fig:GUI} (E)). 

In order to align with Explanatory Debugging Principles 1) Be iterative and 4) Do Not Overwhelm, panels D and E are only revealed when a specific loan application is selected in panel C. While users can provide corrections to the model in panel F, our prototype does not yet integrate this feedback through an online learning process or is able to show any changes to the model based on these corrections to the user on-the-fly. Hence, we do not focus on justifications based on Explanatory Debugging principles dealing with how to integrate feedback. (We show how this user input can be used in off-line experiments in section 4.5.2). 

As the machine learning model for this system, we use a logistic regression model with 26 attributes as explanatory variables. In this model, each attribute has its weights and each applicant has a specific value for each weight. Users can modify the weight for each attribute in each application with slide bars that we will explain in the part of \textbf{Application Details - Suggest changes mode} in this subsection. While we use logistic regression in our system, our interface is applicable to models that use linear functions with the weights for each attribute, such as Naive Bayes classifier or support vector machine. More details of the logistic regression model we used will be explained in Section~\ref{sec:AIModel}.

\begin{figure}[t]
\centering
\includegraphics[width=0.3\linewidth]{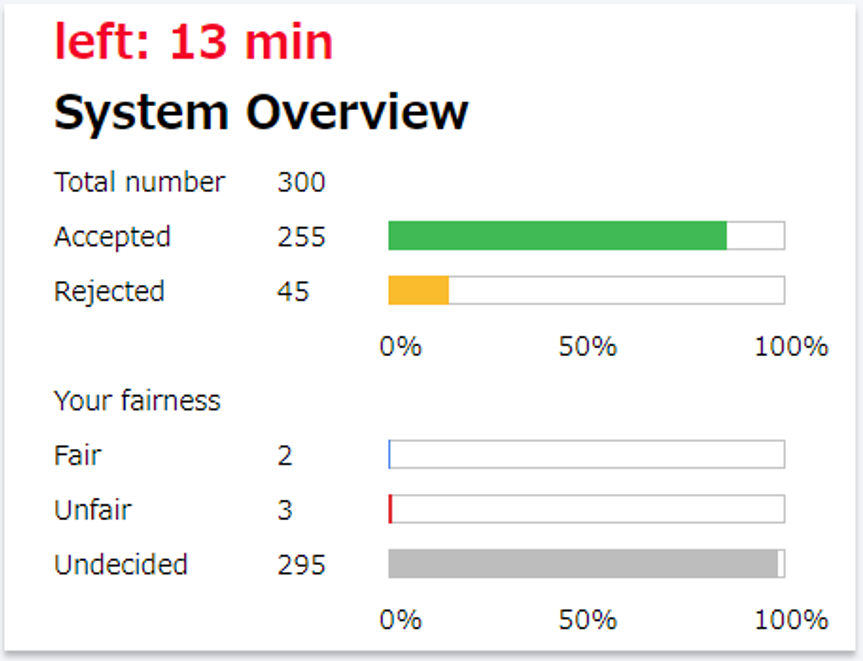}
\caption{\textbf{System Overview} panel.}
\Description[]{}
\label{fig:GUI_A}
\end{figure}

\textbf{System Overview:} Participants in the workshops often used acceptance rates to judge fairness, through defining fairness as \emph{equality of opportunity}, \emph{group fairness} and \emph{individual fairness}. The system-wide information (panel A, Fig.~\ref{fig:GUI_A}) shows the numbers of rejected and accepted applications and the numbers of applications that have been judged fair or unfair by the user. This panel provides a global explanation to the user about decisions that the system has made in terms of accepting or rejecting loan applications, and thus shows information against which to judge fairness. In addition, the panel follows Explanatory Debugging Principle \texttt{1) Be iterative} , as it updates the overview of fairness ratings based on the user feedback.

\begin{figure}[t]
\centering
\includegraphics[width=0.3\linewidth]{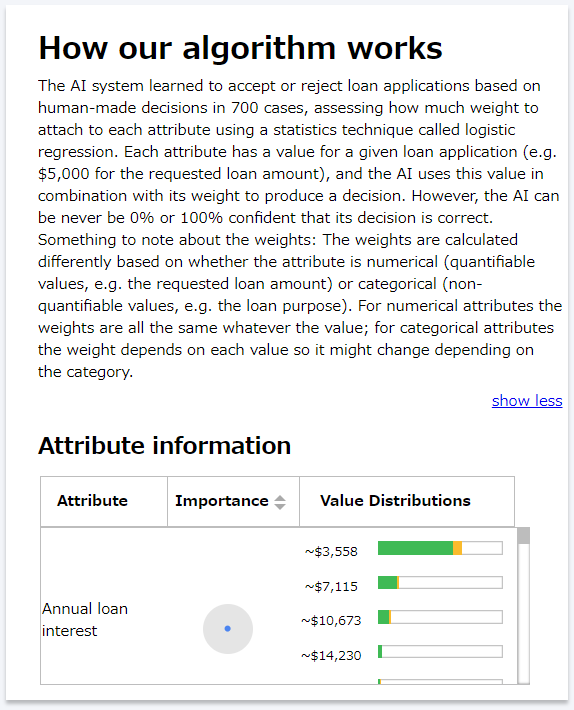}
\caption{\textbf{How Our Algorithm Works} panel.}
\Description[]{}
\label{fig:GUI_B}
\end{figure}

\textbf{How Our Algorithm Works:} The design of this panel was based on feedback by participants as they wanted to know about the \emph{reasons} for making the particular decision, and the underlying \emph{model} for the decision-making. In panel B (Fig.~\ref{fig:GUI_B}), an explanation is provided about how the model algorithm works and model attribute information. The model algorithm explanation aims to explain in simple terms how a logistic regression model makes decisions, which is the model used in our example but this could be adjusted of a different model is used. To save screen space, the user can collapse or expand this explanation.

Participants wanted to know \emph{what information was used in the model to make decisions}. They often feared that there might be other unknown attributes used in the decision-making, and they wanted full transparency of which attributes influenced the decision. In particular, co-designers were worried that protected attributes, such as \emph{gender}, might be applied, or wondered whether (and how much) the \emph{monthly repayment amount} mattered. They placed great importance on \emph{procedural fairness}, in which protected, nonfactual or incorrect information is not used, and that the decision-making follows a logical and fair procedure. A co-designer stated that ``\textit{unfair is being stereotyped, you've got a certain postcode so you must be from a certain community} (UK101).'' 

They stated that knowing the \emph{importance} and \emph{value distributions} can give clues as to any potential biases in the decision-making. Participants were interested in the \emph{attribute weighting} in the model. We showed this as the attribute absolute weight, represented as the size of a circle, and a bar chart of the distribution of attribute values with respect to the application decision. The attributes can be sorted based on their importance. For the value distribution, we divided the minimum to maximum values of continuous variables into five equal parts and showed the percentages of rejected and accepted applications categorized into each part as a stacked bar graph. We represented the percentage of the accepted applications with green, and that of rejected with yellow. The sum of the values in stacked bar graphs for each attribute is 100\%. 

In designing this panel we also took note of Explanatory Debugging principles \texttt{2) be sound, and 3) be complete}. To instantiate 2) be sound, we worked with workshop participants and the research team to develop a simple way of explaining the logistic regression algorithm used in the decision-making while being faithful to the model.  To address 3) be complete, we listed \emph{all} of the attributes that the algorithm is using.

\begin{figure}[t]
\centering
\includegraphics[width=0.3\linewidth]{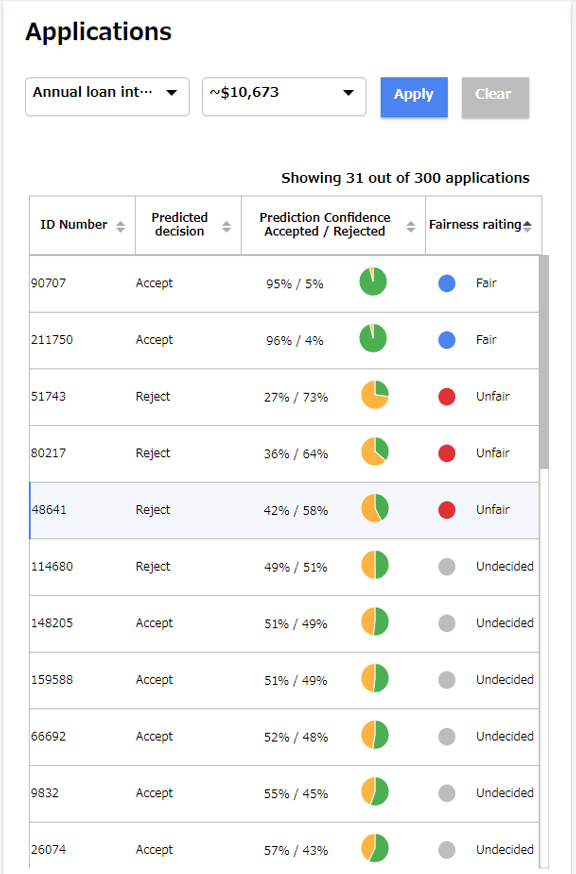}
\caption{\textbf{Application Overview} panel.}
\Description[]{}
\label{fig:GUI_C}
\end{figure}

\textbf{Applications Overview:} In designing this panel, we reflected workshop participants' requirements of wanting to know \emph{how sure the algorithm was in accepting or rejecting the application}, or how close the decision was to being accepted or rejected. This implies that they are interested in the \emph{decision boundaries} of the algorithm, which we represented as a "prediction confidence" percentage. In panel C, (Fig.~\ref{fig:GUI_C}), applications are shown with their ID, the results of the model's prediction (i.e., reject or accept), the confidence of decision represented by percentages as well as a pie chart, and markup by the user of whether they considered the application to be fair or unfair. This percentage was calculated as the probability that an application was accepted. The probability is the value resulting from a sigmoid function where the variable is a linear utility function. The utility function is the inner product between weights for attributes and the values for the attributes plus a constant term. The weights and the constant term were calculated through optimization of the cost function for the logistic regression. When the percentage is more than 50\%, the application is considered accepted.
 
Users can interact with this overview in a number of ways: they can filter applications based on attribute values, or they can sort the application list on ID number, predicted decisions, confidence, and the fairness evaluation by the participant.

Showing the prediction and its confidence instantiates Explanatory Debugging principle \texttt{2) be sound}, and by its simplicity also \texttt{4) do not overwhelm}. Highlighting a specific application brings up more detail about it, shown in panel D and E, and thus follows Explanatory Debugging principles \texttt{1) Be iterative and 4) Do Not Overwhelm}.

\def\gscale{0.6}
\begin{figure*}[t]
\begin{tabular}{cc}

    \begin{minipage}[t]{0.5\hsize}
    \begin{center}
        \includegraphics[scale=0.6]{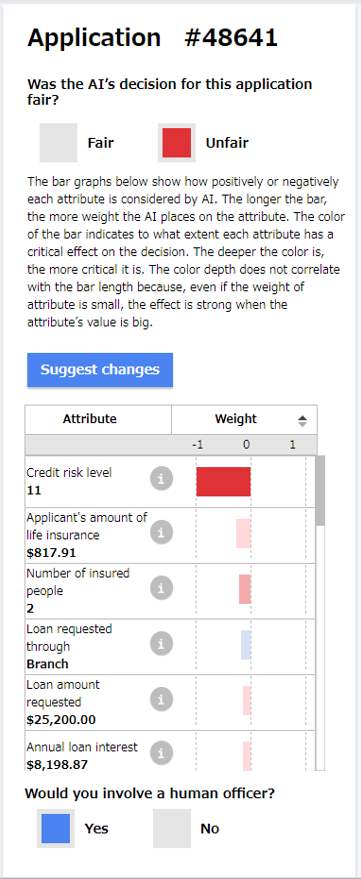}
        \subcaption{The panel before clicking ''suggest changes'' button.}
        \label{fig:GUI_D}
    \end{center}
    \end{minipage}  
    
    \begin{minipage}[t]{0.5\hsize}
    \begin{center}
        \includegraphics[scale=0.3]{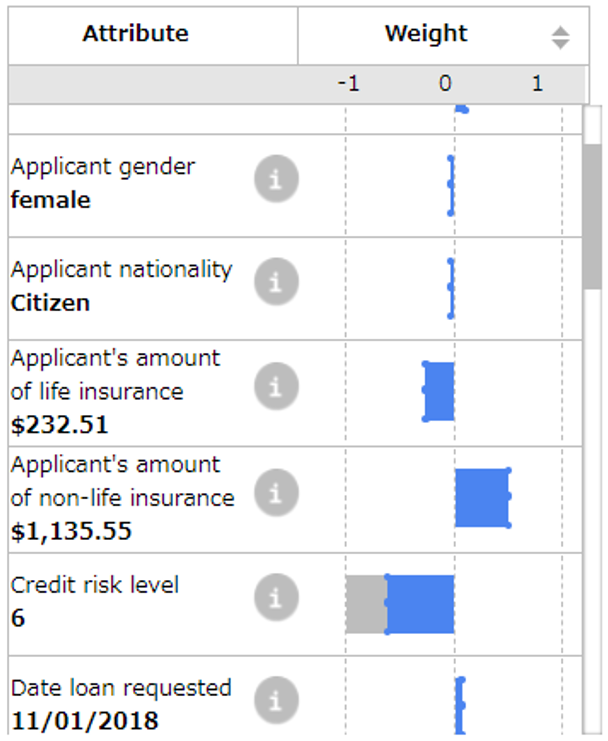}
        \subcaption{The bar charts in the panel after clicking ''suggest changes'' button. Users can directly adjust the weight for each attribute by dragging the blue bar.}
        \label{fig:GUI_F}
    \end{center}
    \end{minipage} \\
\end{tabular}
\caption{\textbf{Application Details} panel.}
\label{fig:GUI_DF}
\end{figure*}

\textbf{Application Details - View mode:} Users can check the details of each application in panel D ( Fig.~\ref{fig:GUI_D}). The user can mark up each application as fair or unfair, which is then also shown in the Applications Overview, panel C. The main part of the Application Details panel is a bar chart, to reflect the insights from the workshops.

Participants often looked at lots of different attributes such as \emph{annual income},\emph{loan amount}, and \emph{employment status} to determine whether the loan was affordable. In addition, they frequently looked in detail at the \emph{postcode/neigborhood}, \emph{purpose of loan}, \emph{residence status}, and \emph{date of birth} attributes that were given as part of the loan application to assess whether a loan decision was fair. We therefore showed each attribute and its value in this panel.

Participants in the co-design workshops wanted to know \emph{how important each attribute was to the decision}, and \emph{what attribute values had positive or negative effects on the outcome}. Alongside this, they requested details about \emph{which attributes had the most critical influence on the outcome}. Consequently, the bar chart shows the weight for each attribute and the criticality for each attribute value. Both show the extent of influence an attribute has on the application's prediction. An attribute's weight is derived from the cost function of the logistic regression. The criticality is the product between the weight for an attribute and the chosen application's value for the attribute. The weight for each attribute is shown in the bar chart as the direction and length of each bar, and criticality (i.e., to what degree each value has an effect on the final decision) with saturation of colors; the more saturated, the higher the effect on the decision (red represents negative effects, blue a positive effect). The bar graphs can be sorted based on the weight. Additional information about the provenance of each attribute value is displayed. The user can also mark up whether the application might need human input.

Provenance information was provided by hovering over an attribute information icon because many participants were also interested in but \emph{unclear about the credit score} and its role in the decision-making, for example: ``UK105: I don’t really know what a credit score of 921 is.''

Finally, the user can also suggest that the decision needs human input. This followed discussions during the workshops indicating that while they thought AI could make decisions fairer by \emph{reducing human bias}, they also stated that some decisions might require some form of \emph{human judgement}. This is because co-designers considered there are some 'gray' areas which require a human to sort out, for example:''\textit{...AI is quite black and white and it doesn’t understand different nuances and biases and I think to make something fair or not fair, you kind of need a person} (UK102).'' 

In designing this panel, we also followed Explanatory Debugging principles \texttt{1) be iterative 2) be sound and 3) be complete}. 

\begin{figure}[t]
\centering
\includegraphics[width=0.3\linewidth]{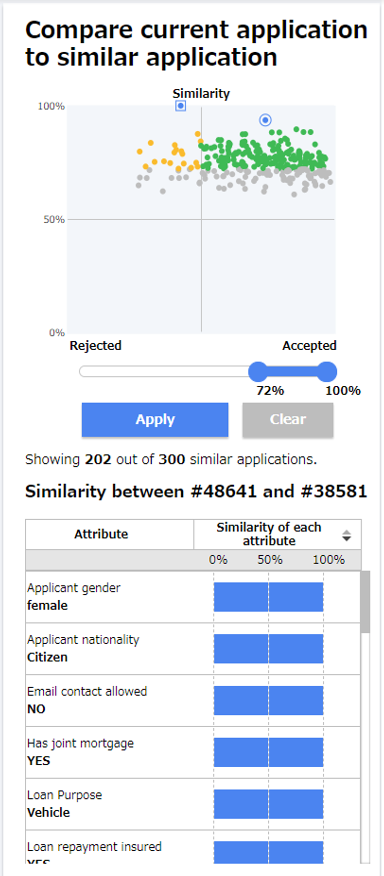}
\caption{\textbf{Compare Current Application to Similar Application} panel.}
\Description[]{}
\label{fig:GUI_E}
\end{figure}

\textbf{Application Details - Suggest changes mode:}
The design of this mode addresses workshop participants' requirement to \emph{make changes to the attribute importance and weighting}. After entering the 'suggest changes' mode (Fig.~\ref{fig:GUI_F}), participants can change the weights for each attribute by dragging and dropping the bar graph.
 
 This mode also follows Explanatory Debugging principle \texttt{5) be actionable}, by allowing user feedback on explanations. They also partly align with Explanatory Debugging principle \texttt{6) be reversible}, as any edits can be cancelled or overwritten.

\textbf{Compare Current Application to Similar Application:} Panel E (Fig.~\ref{fig:GUI_E}) supports users to further delve into evaluating the fairness of the decisions and the model. Workshop participants often wanted to compare applications in some way to do this. In particular, participants stated the need for \emph{knowing the number of similar applications with the same or different outcomes}, being able to \emph{compare the details of similar applications} and \emph{compare similar applications with different outcomes}, especially if they crossed the decision boundary. To satisfy these requirements, we designed Panel E to allow the user to compare the currently selected application with other applications. Users see a scatterplot of applications based on their similarity to the currently selected application in panel C and their prediction confidence. Here, we calculated similarity between two applications as the average value of similarities of all attribute values. Applications are shown in green if they were accepted, and yellow if rejected. Participants can filter the applications based on the similarity between each application and the selected application with the slide bar below the scatter plot. Applications whose similarity is out of the range are grayed out and become unselectable. To further compare a particular application, the user can select it in the scatterplot. This then shows a comparison of the application details, and the respective attribute similarities. 

This panel again was designed according to Explanatory Debugging principles \texttt{1) be iterative 2) be sound and 3) be complete}.
 
\section{An Empirical Study of Using the Prototype}
 
In order to investigate how end-users assess whether an AI system is fair or unfair, using a human-in-the-loop interface (RQ2), whether they can fix fairness "bugs" through their feedback (RQ3) and to explore cultural dimensions to fairness assessments and use of the prototype (RQ4), we set up an online study that asked a large pool of participants to interact with an AI model through the prototype. Due to COVID-19 restrictions, we were unable to conduct face-to-face observational sessions.

\subsection{Participants} 
We recruited 388 participants (129 female, 256 male, 2 Other and 1 preferred not to say) through Prolific\footnote{https://www.prolific.co/}, an online research platform, and paid them £3.50 for an expected 30-minute session. Again, as with the co-design workshops, we did not require any technical or domain knowledge of fairness in loan applications, however, more than a half of our participants had some programming experience (26.8\%), familiarity with AI, machine learning or statistics (31.7\%), or both (18.6\%), and 146 participants had at least a Bachelor degree. These rather educated and tech-savvy participant profiles are in line with other crowd-sourcing studies \cite{ross_who_2010}. 

We recruited globally but the majority of participants came from Europe (345 people). The remainder came from Latin America (18), North America (11), Africa (10), Asia (2) and the Middle East (2). To investigate cultural aspects, we extended participants' data with Hofstede's country scores for each Cultural Dimension based their reported country of birth and country of residence, as we explain in section \ref{sec:CS_DataAnalysis}.
As we will explain in the section, the participants' cultural scores vary a lot in each cultural dimension, even though they are drawn mainly from Europe.

The study was approved by the City, University of London Computer Science Research Ethics Committee and all participants had to be over 18 years of age and give their informed consent. 

\subsection{The AI model}
\label{sec:AIModel}

Instead of using an open-source dataset, the AI system we developed was based on an anonymized loan decision dataset we obtained from a project partner that provides financial services. This dataset contains decisions made on 1000 loan applications and has 35 attributes including the label of whether the loan application was accepted or rejected. These attributes include demographic information of the applicant (age, gender, nationality, etc), financial information (household income, insurance, etc), loan information (amount of loan requested, purpose of loan, loan duration, monthly payments, etc), as well as some information of their financial and banking history (years of service with the bank, etc). There were also some attributes that related to internal bank procedures, such as a money laundering check and a credit score developed by the bank. We removed attributes that lacked more than 100 (10\%) cases, or where multiple attributes had similar values. As a result, we used 26 attributes in our AI model. We then randomly split the dataset into 70\% training data and 30\% test data, and trained a logistic regression model. To train the model, we applied a limited-memory BFGS as an optimization algorithm and an L2 regularizer with default settings. The balanced accuracy of the resulting model was 0.618 for the test data.

Note that the model we developed was unfair; this unfairness in the model occurred naturally in the process of training the model and is one way how many AI model end up with fairness issues without explicit bias in the data. The particular attribute that caused unfairness was Nationality. This attribute has two values: 'citizen' and 'foreign', where citizen means the customer's nationality is the same as the bank's location, and foreign means that the applicant is from a different country. In our AI model, foreigners tended to be rejected more frequently than citizens. Recall that disparate impact, a widely used fairness metric, is calculated as the ratio of the percentage of the accepted people in the group of interest and the percentage in the other groups; anything below 0.8 is considered unfair. Using disparate impact as a fairness metric, our model is unfair (0.718) on  Nationality. This is in contrast to the original dataset, where disparate impact of Nationality is 0.854, i.e. is considered fair.

\subsection{Procedure}

The study session consisted of four phases: pre-questionnaire, tutorial, use of the interface to assess fairness, and a post-questionnaire. First, in the pre-questionnaire (2.5 minutes), we asked participants for background information as well as to rate their general perception of AI fairness on a 7-point Likert scale. After the pre-questionnaire, we presented a tutorial page which explained the organisation and functionality of the interface to familiarise them with its use (5 minutes). 

For the main part of the session, we asked participants to interact with the interface for 20 minutes. We did not give them specific task instructions; instead we asked them to consider whether the system makes fair decisions on loan applications by looking at applications it has made decisions on, interacting with the interface to give them information on applications that might help them to make this assessment, and also to make suggestions that might help to make the system better in the future. During this part, we presented the 300 applications in the test data along with the predicted label of whether to accept (predicted probability is more than 50\%), or reject (predicted probability is the same as or less than 50\%). Participants were free to use the interface as they wanted; after that time, participants were moved along to complete the post-questionnaire. 

In the post-questionnaire (2.5 minutes), we asked again about their AI fairness rating but this time based on their experience of the AI system making decisions on loan applications. We then asked them to describe in their own words what strategies they used to assess the fairness of the system, any systematic fairness issues they had spotted, and their views on suggesting changes and addressing fairness. We then finished the session by asking them to rate their task load using the NASA-TLX questionnaire \cite{hart_development_1988}. All questionnaires are available in the electronic appendices.

\subsection{Data analysis}
\label{sec:CS_DataAnalysis}
We did not integrate explicit attention checks during the procedure so as not to disturb their interactions with the prototype. However, we removed 155 participants from the data analysis who completed the study in an infeasibly short time, who failed to interact with the prototype interface, or who did not complete the associated questionnaires. This resulted in 388 participants whose data we analyzed.

In order to analyze participants' assessments and interactions with the system we drew on a rich tapestry of data using a mixed-methods approach. First, we gathered fairness ratings from the pre- and post-questionnaires, and responses to the NASA-TLX questions. We also logged interactions between the participants and our system, such as clicks and settings of interface elements, applications marked as Fair or Unfair, and the details of suggested changes for further analysis. All these measures lend themselves to quantitative analysis, and where appropriate, we conducted statistical non-parametric tests. Non-parametric tests are less susceptible to imbalanced groups and data which are not normally distributed which makes them appropriate to our data. We also analyzed participants' post-questionnaire responses qualitatively using thematic analysis \cite{braun_using_2006} to supplement and contextualize our quantitative findings.

To investigate the impact of any suggested changes by participants on fairness, we developed a new model off-line in which we  aggregated the suggested weights by the participants. To do so, we first averaged the weight value for each attribute suggested by the participants on an application. We then applied the suggested weights as the weights for the attribute of the application and calculated the confidence based on the weights, keeping the original weights if no changes were suggested for an application. Once we had obtained these new weights, we calculated the disparate impact on nationality as a fairness metric for each of them. 

\textbf{Cultural Dimensions Data Analysis: }We paid special attention to cultural dimension with respect to interactions with the prototype. As we described in Section~\ref{sec:CulturalAspects}, the cultural scores should not be applied on the individual level. Instead, we applied the following procedure. First, we used the participant's country of birth or residence to look up the respective CD scores. We identified the country of residence of 354 participants (91.2\%) from the information registered on Prolific. For 23 participants (5.9\%) in those who had not registered the country of residence, we identified the country of birth registered on Prolific as their country. And for 10 participants (2.6\%) who had not registered any country information on Prolific, we used country of residence registered in the pre-questionnaire as their countries. For only 1 participant (0.3\%) who had not registered the above country information, we use the information of nationality written in our pre-questionnaire as her/his country. If CD scores were not available for a country, we calculated the CD score as the average CD score of countries next to the country. There was only one country (Afghanistan) whose CD score is unavailable.

We then assigned each participant into High and Low groups for each CD dimension, e.g. Power Distance High and Power Distance Low. To make that assignment, we calculated the mean score for each of Hofstede's cultural dimensions based on the country-to-country scores in the publicly available dataset\footnote{https://geerthofstede.com/research-and-vsm/dimension-data-matrix/} and then based the assignment on whether their country-based score was higher or lower than the mean score of the dimension. The mean scores are as follows:  59.33 for Power Distance, 45.17 for Individualism, 49.27 for Masculinity,  67.64 for Uncertainty Avoidance, 45.48 for Long-term Orientation, and 45.43 for Indulgence. For example, a participant from the UK has a Power Distance score of 35; this is lower than the PD mean of 59.33 and hence that participant would be assigned to the Power Distance Low (PD-L) group.  As shown in Table~\ref{tb:Hofstede's Scores}, the participants' cultural scores vary a lot in each cultural dimension, even though they are drawn mainly from Europe, allowing cross-cultural comparison.

To investigate the impact of suggested changes to the AI model based on cultural dimensions, we aggregated the weights as detailed previously for each of the High and Low score groups in the six dimensions.

\begin{table*}
\caption{Average and standard deviation of Hofstede's Cultural Dimensions scores. 
The first row shows the mean values of all country-based CD scores. 
The second row shows the distributions of CD scores for all participants. The third to eighth rows show that the distributions of scores of participants' countries in each region respectively.}
\small
\scalebox{0.87}{
\begin{tabular}{l|c|c|c|c|c|c|c}
\toprule

&& PD & IDV & MSC & UA & LTO & IDG \\
& N & M (SD) & M (SD) & M (SD) & M (SD) & M (SD) & M (SD) \\
\midrule
All countries & - & 59.33 & 45.17 & 49.27 & 67.64 & 45.48 & 45.43 \\
\midrule
All participants & 388 & 57.58 (12.98) & 57.05 (20.38) & 54.48 (17.37) & 82.29 (21.92) & 43.02 (13.88) & 41.83 (18.46) \\
From Africa & 10 & 51.1 (6.3) 	&61	(12.0) 	& 62.7	 (0.9) 	& 49.1	 (0.3) 	& 34	 (0.0) &	60.9	(6.3) \\
From Asia & 2 & 55.5	 (1.5) 	& 37	 (9.0) &	71	 (24.0) &	78.5 (13.5) &	60	 (28.0) &	31	 (11.0) \\
From Europe & 345 & 57.60 (11.83) & 57.65 (19.95) & 53.70 (17.82) & 84.37 (21.54) & 44.48 (13.57) & 37.98 (13.86)\\
From Latin America & 18 & 77.222	 (8.87) &	30.111	 (4.43) &	63.722	(12.97) &	82.667	 (1.49) &	24.717	 (2.38) &	92.074	 (11.75) \\
From North America & 11 & 39.273	 (0.45) &	83	 (4.9) &	54.727	 (4.45) &	47.455	 (0.89) &	33.228	 (4.55) &	68.18	 (0.14) \\
From Middle East & 2 & 13.0 (0.0) & 54.0 (0.0) & 47.0 (0.0) & 81.0 (0.0) & 38.0 (0.0) & 24.0 (0.0) \\
\bottomrule
\end{tabular}
}
\label{tb:Hofstede's Scores}
\end{table*}
\subsection{Results and Findings}

\subsubsection{How participants assessed fairness (RQ2)}

Overall, interactions with the interface were viewed positively by participants. From the post-questionnaires, we found that the options and functionalities provided within the interface appeared to be sufficient for participants, as over 75\% of responses stated that there was nothing else that they wanted to see or do with the system. Many participants also mentioned that they liked the ability to suggest changes for an application to help make fairer decisions. 

\begin{figure*}[t]
 \centering
 \includegraphics[width=\linewidth]{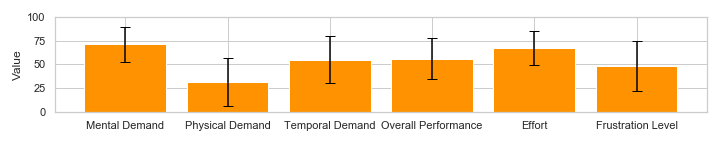}
 \caption{Mean NASA-TLX ratings on each taskload aspect across all participants. Low ratings are better. Error bars indicate standard deviations.} 
 \label{fig:NASA-TLX}
 \end{figure*}

These findings are reinforced by the result of the NASA-TLX questionnaire (Fig. \ref{fig:NASA-TLX}) where participants rated their taskload. Mean ratings for temporal demand, performance and frustration are around 50 which indicates a medium load, while physical demand is low. Mental demand and effort required is relatively high, probably because the interfaces encourages reflection, thinking and exploration which might increase the load in these respects. 

We then investigated how participants made fairness assessments. In the post-questionnaire, participants noted that they were mainly looking at specific decisions (14\%), and over 20\% of participants stated that they relied on the graphs and stats provided in the user interface, such as confidence and value distributions. 
To carry out these assessments, they interacted with the prototype; the mean value that participants clicked UI components during the main task was 142.16 (SD=86.42). Participants frequently sorted on Confidence (33.7\%) and Predicted Decision (28.3\%) out of the number of clicks on all functions (ID Name, Confidence, Fairness rating, and Predicted Decision, 890 clicks in total).
At the same time, 250 (64.4\%) participants used the comparing function (Fig.~\ref{fig:GUI} (E)) and 67 (17.3\%) participants compared individual applications out of 388 participants. These interactions suggest that participants might be using this information to identify applications to further explore in terms of fairness, for example, by focusing on applications close to the decision boundary, looking at applications that were rejected perhaps unfairly, and contrasting applications that on the surface looked similar but that had contrasting outcomes.

Our next step was to explore how many individual application decisions they judged as fair or unfair. Across all participants, a mean of 27.8 applications (SD = 29.2) were judged as fair and a mean of 6.80  applications (SD = 7.86) as unfair, out of the 300 given to them. From this, we calculated the \emph{unfairness ratio} as the ratio of applications judged unfair over all fair and unfair judgement made; the mean was 0.208 (SD = 0.164) across all participants. We then investigated how the unfairness ratio was related to each participants' fairness rating in the post-questionnaire. To do this, we conducted a Pearson's correlation analysis between the AI fairness rating of each participant in the post-questionnaire and their unfairness ratio. We found that there is a highly significant but weak negative relationship (r = -0.221, p <0.001), showing that as the unfairness ratio increases, their perception of the AI system's fairness decreases.

Next, we investigated if there were any patterns in how participants judged the fairness of each application using the prototype. The responses in the post-questionnaire showed that participants assessed fairness of the AI system by using \emph{affordability} (mentioned by 26\% of participants in the post-questionnaire), \emph{equality of opportunity} (16\%), and \emph{individual fairness} (15\%). 

\begin{figure*}[t]
\centering
\includegraphics[width=\linewidth]{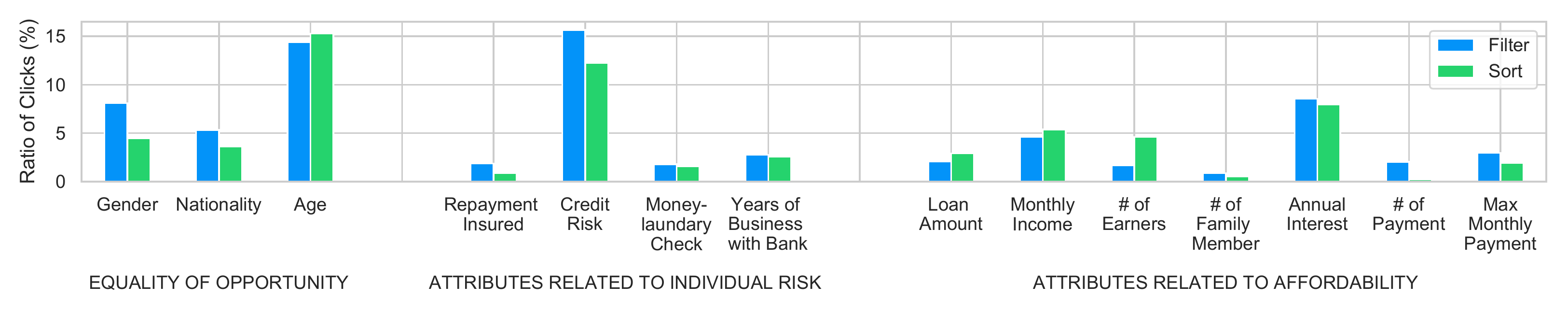}
\includegraphics[width=\linewidth]{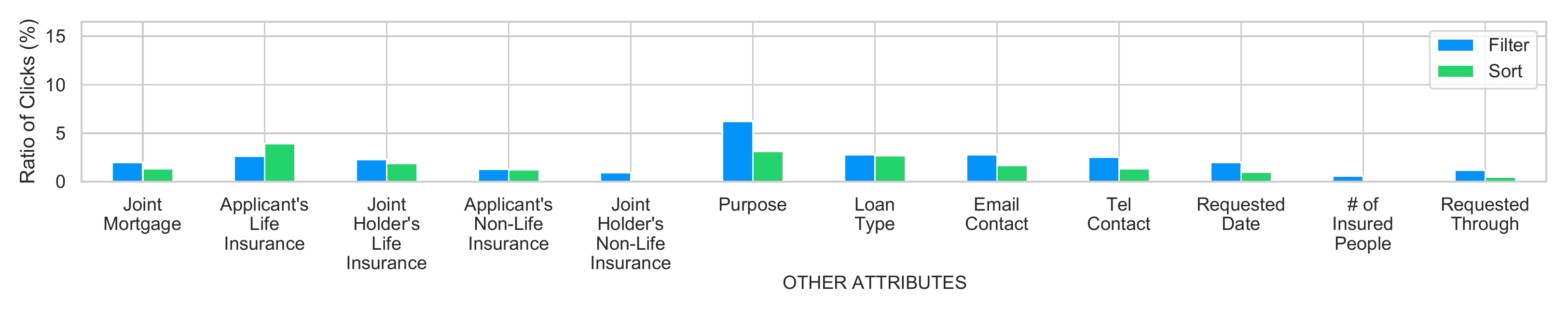}
\caption{The percentage of the number of clicks for each attribute out of the total number of uses of the filter and sort functions respectively.}
\Description[Bar charts showing the ratios of attributes used in the filtering and sorting functions.]{The bar charts show that overall users tended to filter, rather than sort attributes}
\label{fig:sorting _func_ratio}
\end{figure*}

To check these criteria, they focused on particular attributes which is substantiated by our interaction logs (Fig.~\ref{fig:sorting _func_ratio}). For affordability, many participants chose to filter or sort on attributes that relate to this criteria, such as the annual interest, monthly income, number of earners, and loan amount.  A second strategy, checking for equality of opportunity, involved sensitive attributes such as age, gender, and nationality which they frequently filtered and sorted on. Finally, assessing the individual risk of an applicant was carried out using attributes such as credit risk and years of business with the bank, and whether the loan repayment was insured. 

However, participants also interacted with the prototype to investigate a large range of other attributes, especially the purpose of the loan and the loan type, to explore their impact on loan decision and hence fairness of these decisions. This shows that our prototype can support flexible fairness assessments because it allows a range of fairness criteria to be employed and associated attributes to be investigated.

\textbf{Cultural Dimensions (RQ4): }
We did not find any strong patterns in the NASA-TLX responses which cultural dimensions could explain. This indicates that cultural dimensions have no significant effect on the usability of this tool.

However, we found strong evidence that the prototype might be used very differently across cultures, and that different fairness criteria might be applied. First, we delved deeper into cultural differences in terms of whether they judged applications as fair or unfair (Table~\ref{tb:num_of_interactions}).  
First, we looked for differences \emph{between} all groups in how many applications were judged fair or unfair, and found that there was a significant difference (Kruskal-Wallis, H = 24.00, p = 0.013) in the number of applications judged unfair. As a result of a post-hoc Steel-Dwass test, we found there are significant differences between UA-H and UA-L groups (p = 0.044), and between IDG-L and UA-L groups (p = 0.028). Then, we looked for differences \emph{within} dimensions in how many applications were judged fair or unfair (Table~\ref{tb:num_of_interactions}).  We found that there were significant differences for fairness perceptions within the Masculinity, Uncertainty Avoidance, and Indulgence dimensions for applications judged as fair, unfair and the unfairness ratio.  Here, participants in the MSC-H group judged significantly fewer application decisions as fair than the MSC-L group (Mann-Whitney U, U = 18786.5, p = 0.023), participants in the UA-H group judged more application decisions as unfair than the UA-L group (U = 9270.5, p <0.001), and finally participants in the IDG-H group judged fewer application decisions as unfair than the IDG-L group (U = 19046.0, p = 0.002). We did not find any significant differences \emph{between groups} for the unfairness ratio (Kruskal-Wallis, H = 15.74, p = 0.151) but we found significant differences in the unfairness ratio \emph{within} groups, where the UA-H group's unfairness ratio was higher than that of UA-L group (Mann-Whitney U, U = 10020.5, p = 0.014), and that of participants in the IDG-H group was less than the IDG-L group (U = 18415.0, p = 0.011). Taken together, these results indicate that Masculinity, Uncertainty Avoidance, and Indulgence dimensions influenced fairness assessments, and that participants with High Masculinity, High Uncertainty Avoidance and Low Indulgence tended towards assessing the system as more unfair than others.

\begin{table*}
\caption{Details of application decisions judged as fair, unfair and the unfairness ratio, for each Cultural Dimension group.}
\small
\scalebox{0.86}{
\begin{tabular}{c|c||c|c|c|c||c|c|c|c||c|c|c|c}
\toprule
&& \multicolumn{4}{c||}{Judged Fair} & \multicolumn{4}{c||}{Judged Unfair} & \multicolumn{4}{c}{Unfairness Ratio} \\ \midrule
& N & M & SD & U & p & M & SD & U & p & M & SD & U & p\\
\midrule
PD-L & 158 & 24.76 & 23.34 & \multirow{2}{*}{16899.0} & \multirow{2}{*}{0.242} & 6.30 & 7.96 & \multirow{2}{*}{16609.0} & \multirow{2}{*}{0.149} & 0.1998 & 0.16146& \multirow{2}{*}{17268.5} & \multirow{2}{*}{0.406}\\
PD-H & 230 & 29.96 & 32.44 &&& 7.14 & 7.79 & & & 0.2135 & 0.1653 &&\\ \midrule
IDV-L & 109 & 27.06 & 25.76 & \multirow{2}{*}{15883.5} & \multirow{2}{*}{0.495} & 6.58 & 7.37 & \multirow{2}{*}{14609.5} & \multirow{2}{*}{0.547} & 0.1986 & 0.1706 & \multirow{2}{*}{14323.0} & \multirow{2}{*}{0.374}\\
IDV-H & 279 & 28.14 & 30.42 &&& 6.88 & 8.06 &&& 0.2116 & 0.1610&&\\ \midrule
MSC-L & 125 & 31.07 & 29.59 & \multirow{2}{*}{18786.5} & \cellcolor{lightgray} & 7.12 & 7.33 & \multirow{2}{*}{17177.5} & \multirow{2}{*}{0.472}  & 0.1916 & 0.15499 & \multirow{2}{*}{15183.0} & \multirow{2}{*}{0.224}\\
MSC-H &263 & 26.30 & 28.88 &&\multirow{-2}{*}{\cellcolor{lightgray}${0.023}$}& 6.64 & 8.11 &&& 0.2157 & 0.1674&&\\\midrule
UA-L & 79 & 23.33 & 20.51 & \multirow{2}{*}{11182.0} & \multirow{2}{*}{0.250} & 4.54 & 4.77 & \multirow{2}{*}{9270.5} & \cellcolor{gray} & 0.1664 & 0.1388 & \multirow{2}{*}{10020.5} & \cellcolor{lightgray}\\
UA-H & 309 & 28.99 & 30.91 &&& 7.37 & 8.38 &&\multirow{-2}{*}{\cellcolor{gray}$<0.001$}& 0.2186 & 0.1680&&\multirow{-2}{*}{\cellcolor{lightgray}$0.014$}\\ \midrule
LTO-L & 248 & 28.75 & 31.32 & \multirow{2}{*}{17706.0} & \multirow{2}{*}{0.745} & 7.04 & 8.88 & \multirow{2}{*}{17110.0} & \multirow{2}{*}{0.814} & 0.2061 & 0.1672 & \multirow{2}{*}{16786.5} & \multirow{2}{*}{0.589}\\
LTO-H & 140 & 26.22 & 24.90 &&& 6.37 & 6.53 &&& 0.2112 & 0.1578&&\\\midrule
IDG-L & 271 & 29.04 & 31.76 & \multirow{2}{*}{16387.5} & \multirow{2}{*}{0.599} & 7.36 & 7.65 & \multirow{2}{*}{19046.0} & \cellcolor{gray} & 0.2208 & 0.1657 & \multirow{2}{*}{18415.0} & \cellcolor{lightgray}\\
IDG-H & 117 & 25.06 & 21.85 &&& 5.50 & 8.22 &&\multirow{-2}{*}{\cellcolor{gray}$0.002$}& 0.1781 & 0.1555&&\multirow{-2}{*}{\cellcolor{lightgray}$0.011$}\\ \bottomrule
\end{tabular}
}
\label{tb:num_of_interactions}
\end{table*}

We then looked at the potential reasons for making these assessments from the post-questionnaire, based on cultural dimensions. More participants in the MSC-H group (34\%) and the UA-H group (34\%) mentioned assessing \emph{affordability} than in the Low score counterparts (26\% and 23\%). This means that High Masculinity and High Uncertainty Avoidance tended to look for 'objective' criteria to determine whether the loan is fair. We did not find a similar reason to distinguish Low Indulgence in our data.

Other cultural dimensions also seemed to play a factor in the reasons for making assessments and how the tool was used. There was a higher percentage of participants who stated that they checked for \emph{equality of opportunity} and \emph{sensitive attributes} in the IDV-L group (20\%) versus IDV-H (12\%), and also LTO-H group (24\%) versus the LTO-L group (15\%). This also seems to be partially borne out by the analysis of the interaction logs. In a quantitative analysis, we found that the LTO-H group interacted more with for maximum monthly payment (U = 15993.5, p = 0.019). Yet more evidence for the impact of cultural dimensions on the use of the prototype comes from the PD-L group which interacted significantly more with sensitive attributes such as gender (U = 19716.5, p = 0.045) and nationality (U = 19868.5, p = 0.020) than the PD-H group, and also with attributes such as number of maximum monthly payment (U = 19512.0, p = 0.025), income contributor (U = 19065.5, p = 0.044), and monthly income (U = 18818.0, p = 0.022). This means that equality of opportunity and attributes indicating this fairness criteria seemed to be especially important to Low Individualism, High Long Term Orientation and Low Power Distance.

\subsubsection{Did Participants improve Fairness? (RQ3)}

Recall that our AI model was unfair with respect to the Nationality attribute; foreigners were discriminated against by rejecting more of their applications. While many participants said in the post-session questionnaires that they could not find any systematic errors in the decision-making, many others interacted with the prototype to indicate possible problems with the AI model. One way to do this is to mark an application as unfair. There were significant correlations between an application's unfairness ratio and where participants had interacted with the Credit risk attribute (r = 0.650, p<0.001) or Years of business with the bank (r = -0.260, p <0.001) within the prototype. 
Further, we also found that the unfairness ratio was significantly higher for applications involving foreigners  (M = 0.243, SD = 0.134) than for citizens (M = 0.159, SD = 0.130) (U = 301.0, p <0.001). When we investigated the distribution of applications marked unfair for accepted and rejected applications by citizens and foreigners out of the total unfair judgement (2637), we found that unfair judgements accounted for 57.6\% for accepted citizens, 20.1\% for rejected citizens, 7.9\% for accepted foreigners, and 14.4\% for rejected foreigners. This means that participants overwhelmingly focused on \emph{accepted} applications than \emph{rejected} ones. Participants seemed to be able to successfully identify the fairness issue in the nationality, and convey this through the prototype.

To address any fairness issues, we gave participants the opportunity to suggest changes by adjusting the attribute weight. We found that 230 participants suggested weight changes, to 3.71 (SD = 7.58) applications on average.

\begin{figure*}
     \begin{center}
     \includegraphics[scale=0.49]{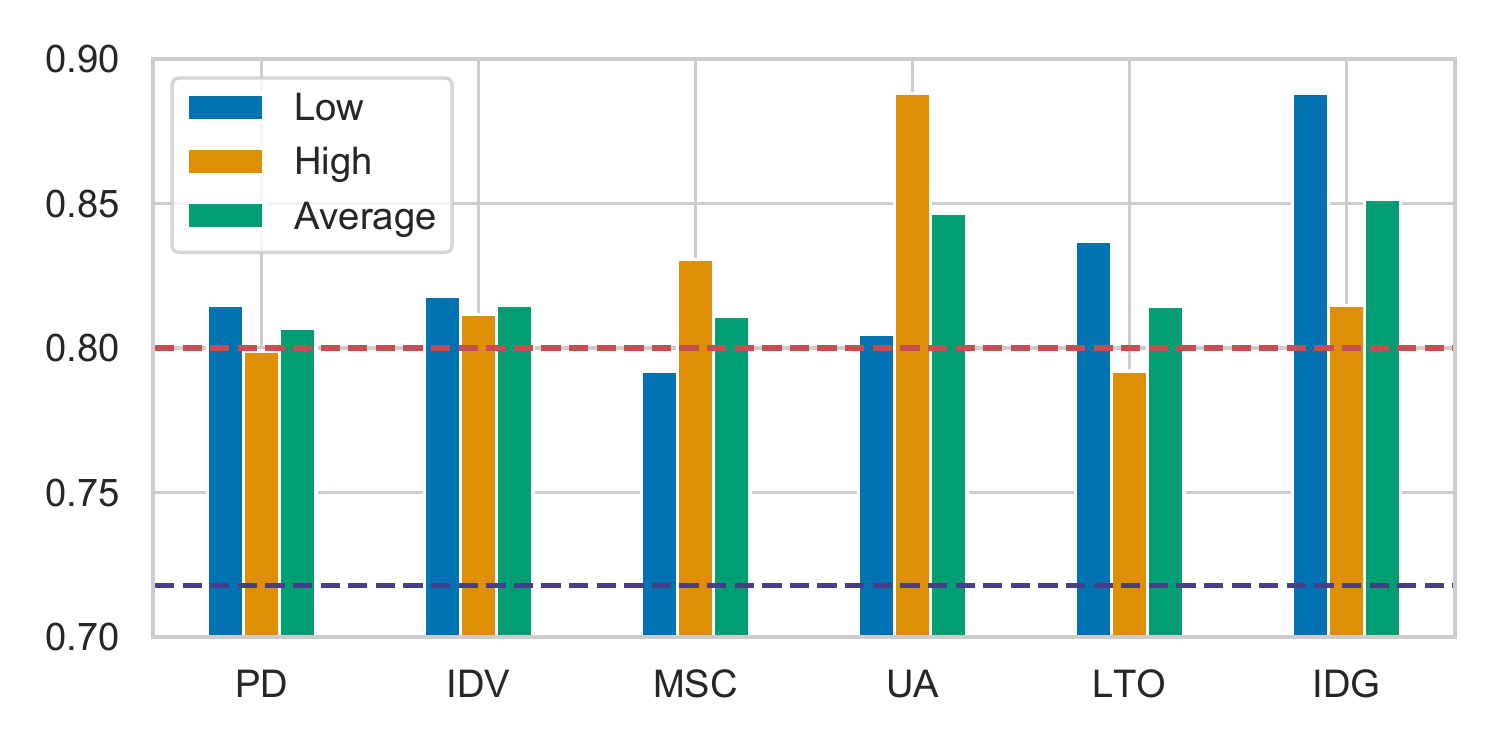}
     \end{center}
\caption{The disparate impact for each Cultural Dimension group. The blue line is the disparate impact of the original results of logistic regression (0.718). The red line is the value above which there is no discrimination (0.80). }
\Description[Clustered bar charts showing the disparate impact of all groups and the average value of each dimension.]{}
\label{fig:DI_BA_newmodel}
\end{figure*}

An important consideration is whether these suggested changes make the AI system fairer, or in fact might make it even worse. Following the approach described in subsection~\ref{sec:CS_DataAnalysis}, we found that across all participants' changes, the value of disparate impact increased from 0.718 to 0.814 based on all participant feedback. We then also explored if individual end-users can improve the mode's fairness. For this, we compared the original DI (0.718) with the DI of a model for each participant who changed weights, trained on their mean values of suggested weights across multiple applications. We found that 115 participants (50\%) increased DI (M = $0.91$, SD =$0.15$) and 115 participants (50\%)decreased DI (M = $0.63$, SD =$0.070$). Taken together, these results show that weight changes suggested by end-users can mitigate discrimination in a human-in-the-loop AI model, however, not all end-users give good feedback. Being unable to 'converge' on a better model is an important limitation for human-in-the-loop learning and explanatory debugging which we will discuss further in section 5.2.

As a follow-up to investigating weight changes by individual participants, we also compared the weights suggested for each attribute between those participants who increased DI and those who decreased DI, using a Mann-Whitney U test. This allowed us to explore which attributes were possibly at the root of the increase or decrease in the DI. Surprisingly, we noted that there was not a significant difference in the suggested weights for Nationality ($p=0.27$). However, we found four attributes had large differences in the suggested weights between these two group: 
Credit risk level ((M, SD): (-0.813, 0.35) for increase, (-1.0, <0.001) for decrease, $U=4652.5$, $p<0.001$ ),
Has joint mortgage ((M, SD): (-0.143, 0.078) for increase, (0.134, 0.0245) for decrease, $U=5679.5$, $p=0.0386$ ), 
Type of loan ((M, SD): (-0.0714, 0.047) for increase, (0.03584, 0.0604) for decrease, $U=2790.0$, $p<0.001$), and
Years of business with the bank ((M, SD): (-0.214, 0.101) for increase, (0.187, 0.0697) for decrease, $U=5783.5$, $p=0.0732$ ). This means that these attributes had a major influence on the fairness of the AI model.

\textbf{Cultural dimensions (RQ4): } We tried to determine if cultural dimensions played a role in making the system fairer. First, we investigated whether cultural dimensions meant they interacted differently with the prototype. We noted that the
LTO-H group suggested weight changes for significantly more applications than the LTO-L group (U = 15258.0, p = 0.0392); otherwise we did not find any differences between either groups or within dimensions.  Second, we also explored cultural differences in making the models fairer (Fig.\ref{fig:DI_BA_newmodel}). We found that all groups made the AI model fairer, but the MSC-L (0.792), LTO-H (0.792) and PD-H (0.799) groups did not make it fair enough. We also noted that IDG-L and UA-H groups had the highest disparate impact score (0.888), which means that they made the AI model very fair. Recall from the previous section, that participants with Low Masculinity judged significantly fewer decisions as unfair than the High Masculinity group, while there were also differences in use of the prototype for High Long Term Orientation and Low Power Distance. This might explain why these dimensions might not made the system fair enough. Also recall that participants with Low Indulgence and High Uncertainty Avoidance judged more decisions as unfair. This could explain why these groups made the system fairest.

\section{Discussion}
\subsection{Including end-users in fairness}
With the development of tools such AI Fairness 360 \cite{Bllamy2019AIF360}, the What-if tool \cite{Wexler2019WhatifTool} and interfaces such as Silva~\cite{Yan2020Silva}, FairSight~\cite{ahn_fairsight:_2019} and FairVis \cite{Cabrera-fairvis-2019}, human-in-the-loop fairness is becoming a reality. However, there is still a shortage of tools that involve end-users in assessing fairness, with most approaches squarely targeted at data scientists and ML experts. 

To the best of our knowledge, our work is the first to consider how end-users could be involved in interactive human-in-the-loop fairness, by providing a prototype that makes decision-making more transparent and allows them to interactively explore decision-making to identify fairness "bugs", inspired by Explanatory Debugging \cite{kulesza_principles_2015} used in interactive machine learning. We found that end-users attended to a wide range of attributes to assess the fairness of a model, often going beyond sensitive attributes such as gender, age, etc. Making model and attribute information transparent through confidence, weights, attribute values helped them hone in on individual applications that might be problematic. They were able to compare these individual applications with other applications to narrow down on potential unfair decisions. Fairness criteria that were often used by end-users included affordability, equality of opportunity and individual fairness. 

In contrast to many other tools, our prototype did not rely on or employ any fairness metrics in assessing fairness. Fairness metrics have attracted some criticisms as it is not clear which one is the right one to choose for a particular context, and that it has to be decided what fairness is a priori. Instead, our approach side-steps this problem; by identifying concrete fair or unfair instances, end-users are able to build up intuitive notions of fairness based explicitly in a given context. 

It could be argued that it is not a good idea to involve end-users in fairness in the first place. Some might suggest that end-users are not technically savvy enough to make these decisions, or that fairness is a human right that should be universally and consistently applied \cite{kirkham_using_2020}. There is also evidence that end-users can be prone to bias \cite{wang_factors_2020} and that their biases could be further replicated in AI. However, this ignores current practices where end-users have been effectively removed from any control over the ways that AI might impact them, and that instead algorithmic design is monopolised by a cadre of ML experts and data scientists. We see interactive human-in-the-loop fairness tools as providing back long-needed agency over AI decision-making to end-users.

To do so, we need to rethink the current AI development process, and particularly processes to ensure responsible AI. The need to directly involve end-users in assessing the fairness of AI systems and making ML models fairer has been emphasized \cite{binns_its_2018, dodge_explaining_2019,holstein_improving_2019, veale_fairness_2018, yan_silva_2020, Yu_DIS20}, and we believe our tool is a step in that direction. While end-users should not be the sole arbiters in developing fair AI, they might provide important input to other stakeholders in the development process. Further work is necessary to reconfigure responsible AI development processes and find suitable stages in which end-users' feedback can be integrated. Recent research \cite{lee2019.WebuildAI} in algorithmic governance suggested a participatory approach to AI design, by building up individual AI belief models and collective aggregation. Our approach might have a place in this framework.

\subsection{Integrating user feedback to fix fairness issues}
Our prototype allowed end-users to make suggestions by changing the attribute weights. This helped in making the AI models fairer, using disparate impact as a fairness metric. Of course, as we pointed out earlier, this might not be the appropriate fairness metric to apply. Further work is needed to investigate how the participants' feedback would fare using other fairness metrics, including the end-users' own criteria.

Our work indicates that it is possible to improve fairness of an AI model using the suggestions that end-users give. Previous work in interactive machine learning \cite{kulesza_principles_2015, stumpf_interacting_2009, fiebrink_human_2011} has already highlighted the value of incorporating user feedback in improving the \emph{accuracy} of AI models, however, allowing end-users to directly influence the \emph{fairness} of an AI system has not received much consideration. Our work showed that across all users, changes to weights improved the fairness of the AI model. However, we also noted that some user input could make fairness worse. This is obviously a concern for human-in-the-loop learning as it is only as good as the input the end-user provides. Ideally, mechanisms will be developed to assess whether end-users give good feedback, or guard against incorporating bad input.

There are possible other ways to incorporate user feedback into these models \cite{yan_silva_2020}. For example, end-users could be allowed to change the predicted labels and train new models based on this adjustment. While is a direct way of obtaining user feedback, there is a risk that resulting models might not be fairer than previous versions as machine learning methods might not overcome inductive bias by incorporating only a few training examples.
Another way is to introduce causal graphs into model development and provide end-users the ability to change causal relationships between attributes. This approach might be more suitable as it also lets end-users understand the detailed mechanisms of machine learning and provide feedback at the same time. Last, showing the impact of adjusting weights to end-users through  fairness metrics might make their feedback more effective.

\subsection{Cultural dimensions}
To our knowledge, cultural dimensions in AI fairness assessment have not been considered, especially applied to the credit scoring or loan application domain. Our results showed that cultural dimensions can explain differences in perceiving, assessing and improving fairness. We found that especially High Masculinity, High Uncertainty Avoidance, and Low Indulgence matter in whether application decisions were assessed as unfair, and also making the AI system fairer.

This has important implications for understanding the contextual nature of fairness. What our results suggest is that fairness perceptions could change based on geographic location, as they underlie cultural dimensions. For example, end-users from countries such as Portugal (UA-H: 104, IDG-L: 33) and Sweden (UA-L: 29, IDG-H: 78) might make very different fairness assessments to each other due to their different respective Masculinity and Indulgence values; based on what we found, Portugese people might judge applications as unfairer than Swedes. However, currently we do not have enough data to confirm this and further studies would be needed to investigate whether there are indeed differences due to relatively small geographical distances. 

Our findings on cultural dimensions have implications for the research, design and use of tools that involve end-users in fairness assessments. Our work cautions against conducting research in AI fairness with restricted user populations. In particular, studies in AI fairness should be extended to include non-WEIRD (Western, Educated, Industrialized, Rich, Democratic) participants \cite{henrich2010most}. In terms of design, tools such as ours seem to be \emph{usable} universally across different cultures. However, they will be used in different ways to express nuanced perceptions of fairness with respect to cultural dimensions. This still begs the question of what to do once these nuanced notions of fairness have been be 'harvested'. We believe it might be possible to develop fair AI models by analyzing and considering these culturally dependent perspectives (and possibly counteract them) in the AI development process.

Our results also extend our understanding of Hofstede's framework of cultural dimensions. We saw in our studies that there were combinations of cultural dimensions that appeared again and again in our findings. This could be explained by how Hofstede's framework was developed: Indulgence was developed relatively recently by refining the Long Term Orientation dimension, while Power Distance might be correlated with Masculinity \cite{hofstede_dimensionalizing_2011}. We look forward to future research to clarify how cultural dimensions might be related. With more data, our understanding of patterns of cultural differences and their impact on fairness assessments could also be refined. 

\subsection{Limitations}
It could be argued that we are conflating \emph{debugging}, \emph{debiasing}, and \emph{mitigation}. We agree that there are subtle but significant differences between these terms. In our view, explanatory debugging underpins the work of individual end-users to find and fix issues, in this case loan application decisions, that run counter to their expectations of fairness. However, this might not systematically expose or rectify bias in the AI model or the data. In addition, there are a number of ways that fairness issues can be robustly mitigated. As we described in section 2.2, there are a number of different mitigation approaches that have been proposed, based on quantitative fairness metrics. Our work has only investigated one technique of mitigation, by using weight adjustments of features, based on subjective user input.

While our prototype allowed interaction by end-users with the AI model to understand how it functioned, it was not a fully interactive ML system. Interactive ML relies on a tight cycle of user-system interaction, where user feedback is integrated into the AI model online, and any changes in the model are communicated back to the user immediately. In our prototype, we used user input on weight changes in an offline experiment, and thus we were not able to show the effects of model updates to end-users. Thus, we were also not able to follow all Explanatory Principles which might have affected the best use of the prototype, and give appropriate fairness assessments. 

We were heavily influenced by Hofstede's work on cultural dimensions. As already mentioned in section 2.4, this framework is not universally accepted, and researchers have instead opted for other kinds of analyses, including country-based comparisons. We believe that cross-cultural studies such as ours are still useful because they draw out similarities and differences in cultural interpretations, rather than geographic locations. While we have focused on cultural aspects across countries, we acknowledge that our participants came mainly from the European region. A larger, more geographically diverse sample would add further validity to our results. 

Our work has investigated fairness in loan applications but there are other domains which our work could be applied to. For example, our prototype could be instantiated for recidivism or medical decision-making, and further studies could extend our understanding of fairness assessments in those domains.

\subsection{Future Work}

Our work showed that AI transparency is very important and intricately entwined with AI fairness. Unless a user can understand how a model works, it will be difficult to assess its fairness. Our results show that end-users were able to understand the model, the attributes, the weights, decision boundaries, confidence, and similarity, if communicated appropriately. However, it has been argued previously that different explanations are needed for end-users, business stakeholders, regulatory bodies, or data scientists \cite{gunning_xaiexplainable_2019}. It is still an open research question how to design appropriate interfaces to different stakeholders so that are understandable and usable to assess fairness. 

We also anticipate that our prototype's functionality could be improved to support fairness assessments. For example, there might be a need to extend comparisons of groups of applications, or to show casual graphs of attributes \cite{yan_silva_2020}. We look forward to further studies that improve the design of interactive human-in-the-loop fairness tools.

Additionally, we recognize that the feedback from end-users might change if there are other constraints on loan decisions, for example,  if there is only a fixed amount that can be lent to customers, or that only a certain number of people can be loaned money from the bank. In such situations, end-users may judge fairness based on other attributes about who 'deserves' the loan, perhaps involving loan amount or sensitive attributes. We hope that there will be further research to extend the knowledge on fairness in these situations.

\section{Conclusion}
In this paper, we have presented an interactive human-in-the-loop prototype that supports end-users in assessing the fairness of an AI system that makes loan decisions, and allows them to feedback changes to the AI model. We employed this prototype in an online study to investigate its use and effects on improving fairness. We paid attention to any cultural aspects in how end-users interacted with the prototype.  We found that:
\begin{itemize}
    \item  End-users assessed fairness using model and attribute information, such as confidence, weights, and attribute values, and by investigating and comparing individual applications using graphical means.
    \item Our prototype allowed them to make suggestions by changing the attribute weights. This helped in making the AI models fairer.
    \item We found that cultural differences explained differences in assessing and improving fairness. The cultural dimensions that seemed to matter most were Masculinity, Uncertainty Avoidance and Indulgence. 
\end{itemize}

Our work suggests future research avenues to study how humans can be involved in fairness. The results we presented form a significant step toward designing for deeper involvement of end-users in AI fairness.

\bibliographystyle{ACM-Reference-Format}
\bibliography{main}

\end{document}